\documentclass[pre,12pt]{revtex4}
\usepackage[usenames]{color}
\usepackage{graphicx}
\usepackage{hhline}
\usepackage{hyperref}
\usepackage{amssymb}
\usepackage{hyperref}
\usepackage{morefloats}
\usepackage{bm}

\usepackage{times}

\usepackage{amsmath}
\usepackage{subfigure}
\usepackage{csquotes}
\usepackage{amsfonts}
\usepackage{amssymb}
\usepackage{graphicx}
\usepackage[usenames]{xcolor}

\usepackage{etoolbox}
\usepackage{setspace}
\usepackage{caption}
\usepackage{ragged2e}

\begin{document}

\title{Hierarchical excluded volume screening\\ in solutions of bottlebrush polymers}
\author{Jaros{\l}aw Paturej${}^{1,2\ast}$, Torsten Kreer${}^{1\ast}$
\\
\normalsize{${}^{1}$Leibniz-Institut f\"ur Polymerforschung Dresden e.V., 01069 Dresden, Germany}\\
\normalsize{${}^{2}$Institute of Physics, University of Szczecin,
70451 Szczecin, Poland}\\
\normalsize{$^\ast$To whom correspondence should be addressed; E-mail:  jpaturej@univ.szczecin.pl}
}

\begin{abstract}
 Polymer bottlebrushes provide intriguing features being relevant both in
nature and in synthetic systems. While their presence in the articular
cartilage optimizes synovial joint lubrication, bottlebrushes offer pathways
for fascinating applications, such as within super-soft elastomers or for drug
delivery. However, the current theoretical understanding lacks completeness,
primarily due to
the complicated interplay of many length
scales.
Herein, we develop an analytical model that
demonstrates
how structural
properties of bottlebrushes depend on the concentration, ranging from dilute
solutions to highly concentrated melts.
The validity of our model is supported by data from extensive molecular dynamics simulation.
We demonstrate that the hierarchical structure of bottlebrushes dictates a sequence of conformational changes
as the solution concentration increases. The effect is mediated by screening of excluded volume interactions at subsequent structural parts of the bottlebrushes.
Our findings provide important
insights that should enable improved customization of novel materials based on the architectural design of polymer bottlebrushes.
 \end{abstract}
\maketitle
\baselineskip24pt

\section*{Introduction}
Bottlebrushes are polymers consisting of linear backbones with densely grafted polymeric side chains \cite{sheiko2,rza,ver,mm}.
Recent progress in polymerization methodologies enables controlled synthesis of these branched macromolecules \cite{beers,li,jha,xia,zhang,ohno,matsuda,bielawski,lin}.
Attaching side chains to the backbone  leads to significant
correlation of the backbone monomers due to the steric repulsion between the
side chains. The variation of architecture, i.e., length and grafting density
of side chains, allows for systematic tuning of molecular
conformation and physical properties of bottlebrushes \cite{schmidt}.
These unique features make bottlebrushes candidates for diverse
applications including super-soft elastomers \cite{pakula,supersoft}, drug
delivery agents~\cite{drug},  molecular sensors \cite{sensors},  stimuli-responsive \cite{stimuli,peng,qzhang} and protective  surfaces \cite{protective}, emulsifiers \cite{emul}, lubricants \cite{lubricant}, porous \cite{porous}, self-assembled \cite{runge,noel}, and thermoplastic \cite{zhangj} materials, photolithography \cite{photo},  ionic transport \cite{ionic}, photonics \cite{photonics}, and energy storage \cite{energy}.
In addition, bottlebrush macromolecules are of great importance in biology,
since a myriad of bottlebrush-like glycoproteins and proteoglycans regulate
crucial functions in the human body including  clearance of lungs \cite{lung},  joint lubrication \cite{joint}, and  cell protection \cite{cell}.

Given that the properties of bottlebrushes are very distinct in comparison to linear polymers, they have been an active field of exploration for
numerous theoretical \cite{bir,borisov,rou,fred,dane,potemkin,ywang,as,ny,rc,feuz,paturej_sci,cao1},  experimental
\cite{supersoft,winter2,lec,rath,vlas,verduzco,boli,feuz2,gcheng,kfisher,bzhang,wliu,carr,kik,yo1,yo2}, and numerical investigations \cite{rou,ten1,ten2,ten3,elli,yet,hsu-str,hsu,feuz2,paturej,paturej_sci,theo,cao1,cao2,paturej_macro}.
A majority
of these studies has focused on basic structural properties in solutions and
in the adsorbed state. Nevertheless, even the most simple question of how the
size of a bottlebrush in dilute solution depends  on  architecture parameters is still a matter of debate.
Several theoretical approaches
have been proposed to address this problem. The main difficulty is the
interplay between  many
length scales characterizing the bottlebrush structure.

Existing analytical theories of bottlebrushes in dilute
solution and under good solvent conditions assume
stretchable or persistent, rod-like configurations of
the backbone at length scales comparable to \cite{bir}
 or larger \cite{fred} than the size of side chains and do not include
 self-consistently the coupling between elasticity of backbone and side chains
 \cite{bir,fred}. The size of the macromolecule predicted from these models scales as $R_0\propto N_{\mbox{\tiny bb}}^{3/5}N_{\mbox{\tiny sc}}^{\alpha}$, with $\alpha$ as low as $9/25$ \cite{bir,fred} or as high as $3/4$ \cite{fred}, where
  $N_{\mbox{\tiny bb}}$ and $N_{\mbox{\tiny sc}}$ respectively are  the
 degrees of polymerization of backbone and side chains.

The concentration dependence of the macromolecular structure of bottlebrushes has received much
less attention.
The model of Birshstein {\em et al.}~\cite{bir} for bottlebrushes in dilute solutions  was used
as a starting point for scaling analysis anticipating a sequence of decays in
the molecular size with increasing concentration, $c$ \cite{borisov}.
 Three different concentration regimes for the size of bottlebrushes are
 predicted in this study: $R_{\rm I}\propto c^{-1/8}$ for semi-dilute solutions,
 $R_{\rm II}\propto c^{-17/56}$ for larger concentrations, and $R_{\rm III}\propto c^{-1/8}$
 for melts.

A thorough understanding of bottlebrush conformations, in particular
under melt conditions,
is vital for the design of novel polymeric materials with superior mechanical properties \cite{supersoft}.
Architecture-induced increase of the bottlebrush persistence length is the key feature underlying
physical properties of melts and elastomers.
Bottlebrushes in a melt behave as flexible filaments,
where side chains reduce
molecular overlap and suppress entanglements among macromolecules \cite{paturej_sci}.
This results in the modification of dynamical and rheological properties
of bulk polymers \cite{pakula,supersoft,neug,grubbs,crowther,bates} and allows for
synthesis of supersoft and superelastic, solvent-free
elastomers, with shear moduli down to $100$~Pa and tensile
strains at break up to $1000\%$ in a solvent free state \cite{supersoft}.
However, before the dynamics of bottlebrush polymers
can be addressed, a solid theory describing their static properties
is highly desirable.

In this work, we present a theoretical approach that allows one to characterize the equilibrium structure of polymers with grafted side chains, covering
dilute, semi-dilute, and highly concentrated solutions.
Our theoretical findings  are compared
to data from extensive molecular dynamics (MD) simulations of a coarse-grained bead-spring
polymer model.
 We identify four concentration regimes for the size of bottlebrushes,
 which gradually establish upon increasing
concentration. They follow from a hierarchical screening of excluded volume effects
on various length scales.
Excluded volume screening first takes place at the largest length scale, along the backbone of the macromolecules,
for concentrations slightly above their overlap concentration. In this regime, the size of the macromolecule scales with concentration as $R_1\propto c^{-1/8}$. Further increase of the
concentration leads to screening at length scales
comparable to the length of  the persistent segment.
The latter results from correlation between backbone monomers due to the steric repulsion among
adjacent side chains.
Here, our theory predicts $R_2\propto c^{-1/4}$.
For even larger concentrations, excluded volume
screening occurs along the interpenetrating side chains of different macromolecules, which yields $R_3\propto c^{-4/13}$. Under melt
conditions, limited interpenetration between bottlebrushes
favors screening of excluded
volume among side chains of the same macromolecule and results in  $R_4\propto c^{-2/5}$.

Our article is organized as follows: In the next section, we discuss
conformational properties of bottlebrushes in dilute solutions using a mean field approach. 
 Then, we present results for the concentration dependence of bottlebrush
conformations obtained from scaling theory
 and compare them to  data of our systematic MD simulations.
 In the last section, we summarize our findings and draw conclusions.
The simulation model and the details of our scaling
analysis are described in the Methods section.

\section*{Results}

\subsection*{Dilute solutions of bottlebrushes}

\begin{figure}[!h]
\center
\includegraphics[scale=0.3]{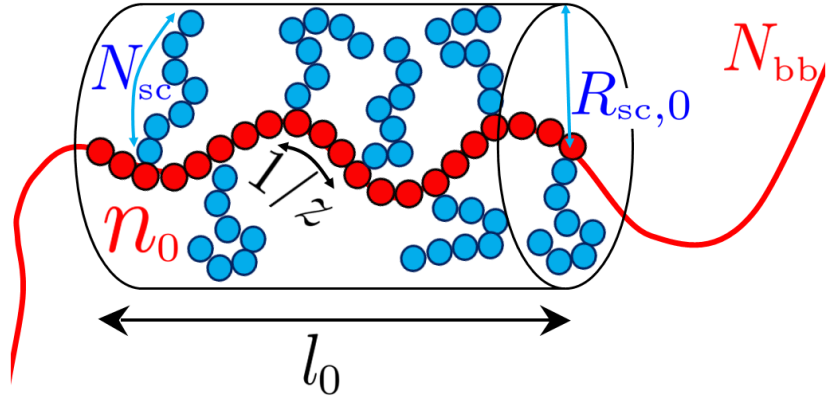}
\caption{{\bf Molecular architecture of a bottlebrush polymer.} The bottlebrush molecule
consists of a backbone with $N_{\mbox{\tiny bb}}$ monomers (red beads) and side chains (blue beads), which are grafted to every $z$-th backbone monomer.
Each side chain is composed of $N_{\mbox{\tiny sc}}$ monomers. The total number of bottlebrush monomers is
$N_{\mbox{\tiny bb}}(1+N_{\mbox{\tiny sc}}z)$. The bottlebrush molecule can be represented as a chain of effective cylindrical
segments. Each segment contains $n_0$ backbone monomers, has a length $l_0$ and a radius $R_{\mbox{\tiny sc},0}$.
}
\label{fig:model}
\end{figure}

The conformation of a bottlebrush macromolecule in dilute solution can be represented as a sequence of cylindrical subsegments, where each cylinder of length $l_0(N_{\mbox{\tiny sc}},z)$
and radius $R_{\mbox{\tiny sc},0}(N_{\mbox{\tiny sc}},z)$ contains $n_0(N_{\mbox{\tiny sc}},z)$ backbone monomers, see
Fig.~\ref{fig:model}.
Here, $N_{\mbox{\tiny sc}}$ is the number of monomers per side chain and $z$ is the grafting density of side chains.
For grafting densities above or
equal to unity ($z\geq 1$), $z$ is the number of grafted side chains per backbone
monomer. Correspondingly, a value of $z< 1$ means that a side chain is grafted to every $1/z$-th backbone monomer.

The cylindrical subsegments can be
interpreted as persistent segments, reflecting the coupling between backbone
monomers due to the steric repulsion between nearby side chains.
The size of a bottlebrush in dilute solution can be estimated from the mean field free energy per cylinder, which reads
\begin{equation}
\label{F.eq}
F\propto\frac{l_0^2}{n_0} + \frac{R_{\mbox{\tiny sc},0}^2}{N_{\mbox{\tiny
      sc}}}n_0z + \frac{n^2_0(1+N_{\mbox{\tiny sc}}z)^2}{l_0R_{\mbox{\tiny sc},0}^2}.
\end{equation}
In the above equation, we have neglected numerical pre-factors and set the excluded volume
parameter, the thermal energy, and the effective monomer size (or Kuhn length)
to unity. Note that backbone and side chains are assumed to be of the same
chemical nature.

The first and the second term in Eq.~(\ref{F.eq}) respectively describe   the
stretching energy of  backbone and  side chains, while the third term accounts for
the overall excluded volume interaction among all monomers within the
cylinder. The latter term  overestimates the excluded volume interaction as it is
commonly done for polymers in good solution \cite{degennes}. However,
previous approaches \cite{rou}, which utilize similar expressions describing the free
energy for the entire macromolecule (and not for a subsegment), invoke an  even larger
overestimation of the excluded volume interaction.

The free energy of Eq.~(\ref{F.eq}) has to be minimized with respect to both $l_0$ and $R_{\mbox{\tiny sc},0}$. The derivatives
$\frac{\partial}{\partial l_0}F=0$ and  $\frac{\partial}{{R_{\mbox{\tiny sc,0}}}}F=0$ yield
\begin{eqnarray}
l_0 &\propto&  n_0R_{\mbox{\tiny sc},0} \sqrt{\frac{z}{N_{\mbox{\tiny sc}}}} \quad\mbox{and}
\label{l.eq}\\
R_{\mbox{\tiny sc},0}  &\propto&  (1+N_{\mbox{\tiny sc}}z)^{2/5}\left(\frac{N_{\mbox{\tiny sc}}}z\right)^{3/10}.
\label{r.eq}
\end{eqnarray}

\begin{figure}[!h]
\center
\includegraphics[scale=0.25]{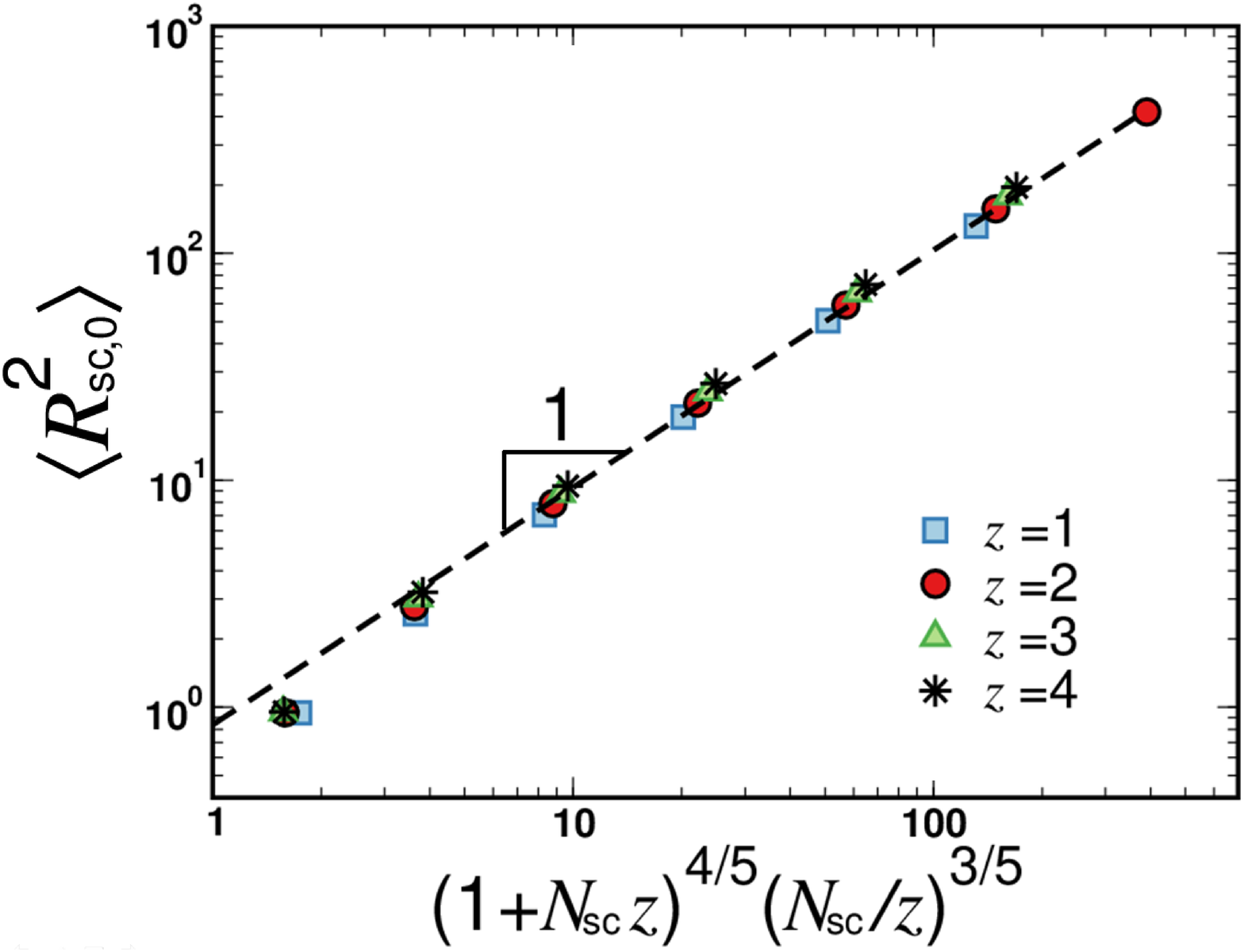}
\caption{{\bf Size of bottlebrush side chains in dilute solution.}
Mean-square end-to-end distance of
side chains, $\langle R^2_{\mbox{\tiny sc},0}\rangle$, in  dilute solution  (concentration $c=10^{-3}$) as
a function of the scaling variable, $(1+N_{\mbox{\tiny sc}}z)^{4/5}(N_{\mbox{\tiny
    sc}}/z)^{3/5}$. 
Data are presented for various grafting densities, $z$, as indicated in the legend, and fixed degree of polymerization of backbone, $N_{\mbox{\tiny bb}}=100$.
The dashed line represents the theoretical prediction of  Eq.~(\ref{r.eq}).
Error bars for all data points are smaller than symbol size.
}
\label{fig:side}
\end{figure}

In Fig.~\ref{fig:side}, we compare the results of our MD simulations for the
mean-square end-to-end distance of side chains, $\langle R^2_{\mbox{\tiny sc},0}
\rangle$, plotted as a function of the scaling variable, $(1+N_{\mbox{\tiny sc}}z)^{4/5}(N_{\mbox{\tiny sc}}/z)^{3/5}$, to the prediction of
Eq.~(\ref{r.eq}). For large enough values of $N_{\mbox{\tiny sc}}$ and $z$ ($N_{\mbox{\tiny sc}}z\!\gtrsim \!8$) the theoretically estimated mean-square size of side chains demonstrates perfect agreement with the simulation data.
Note that in Fig.~\ref{fig:side} and in what follows, all data
are presented in Lennard-Jones units (see Methods section for details).

If grafting density and number of side chain monomers are not extremely large, one can assume a semiflexible backbone inside
the cylinder \footnote{In the following, we do not address the case of ``toroidal" bottlebrushes \cite{fred}, where the backbone becomes locally rod-like at very large values of $N_{\mbox{\tiny sc}}$ and $z$. However, we discuss aspects of this case at the beginning of the Methods section.}.
 Then, $R_{\mbox{\tiny sc},0}$ is the only length scale that
determines the dependence of the persistence length on $N_{\mbox{\tiny sc}}$
and $z$, i.e., $l_0\propto R_{\mbox{\tiny sc},0}$.
This implies that intrinsic rigidity forbids bending of the backbone on length scales smaller than $R_{\mbox{\tiny sc},0}$.
From Eq.~(\ref{l.eq}), we obtain
\begin{equation}
\label{n.eq}
n_0 \propto \sqrt{\frac{N_{\mbox{\tiny sc}}}z},
\end{equation}
such that Eq.~(\ref{r.eq}) leads to
\begin{equation}
l_0\propto  R_{\mbox{\tiny sc},0} \propto (1+N_{\mbox{\tiny sc}}z)^{2/5}n_0^{3/5}.
\label{lsaw}
\end{equation}
The latter equation reveals a self-avoiding walk statistics of the backbone monomers inside
the cylinder, i.e., $l_0\propto n_0^{3/5}$ \footnote{See also discussion above Eq.~(\ref{dsaw}), where we derive this result  using scaling arguments.}.
Correspondingly, the conformation of the  bottlebrush as a whole can be regarded as a
self-avoiding walk of $N_{\mbox{\tiny bb}}/n_0$ cylindrical segments of length $l_0$, where $N_{\mbox{\tiny
    bb}}$ denotes the (total) number of backbone monomers.
 With Eq.~(\ref{lsaw}), the equilibrium size of a bottlebrush in dilute solution reads
 \begin{equation}
\label{rsd.eq}
R_0\propto l_0\left(\frac{N_{\mbox{\tiny bb}}}{n_0}\right)^{3/5}\propto (1+N_{\mbox{\tiny sc}}z)^{2/5}N_{\mbox{\tiny bb}}^{3/5}.
\end{equation}

\begin{figure}[!h]
\center
\includegraphics[scale=0.25]{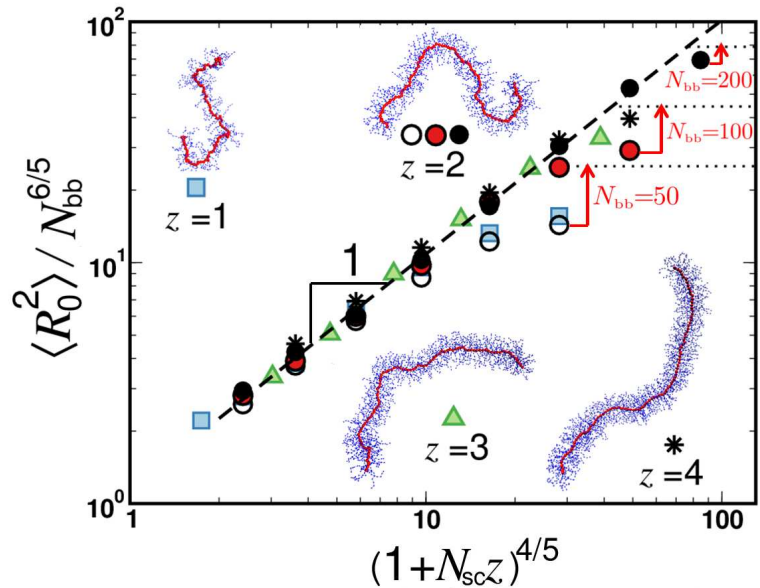}
\caption{{\bf Size of bottlebrushes in dilute solution.}
   Mean-square end-to-end distance,
  $\langle R_0^2\rangle$, of bottlebrushes in  dilute solution (concentration $c=10^{-3}$), normalized by the dependence of
$\langle R_0^2\rangle$ on the backbone's degree of polymerization in good
solvent, $N_{\mbox{\tiny bb}}^{6/5}$, as a function of the scaling variable,
$(1+N_{\mbox{\tiny sc}}z)^{4/5}$. The dashed line represents the theoretical
scaling law [see Eq.~(\ref{rsd.eq})].
The number of side chains grafted per backbone monomer is $z=1$ (squares), $2$ (circles),
 $3$ (triangles), and $4$ (stars).
 The degree of polymerization of
backbones is fixed to $N_{\mbox{\tiny bb}}=100$, apart of the data
  for $z=2$, where $N_{\mbox{\tiny bb}}=50$ (empty circles), $100$ (red circles), and $200$ (black circles).
For data with $z=2$, red arrows indicate the maximum square end-to-end
distance corresponding to the fully extended backbone (horizontal dotted lines).
 Snapshots display conformations of macromolecules with $N_{\mbox{\tiny bb}}=100$ for various values of $z$, as denoted. For all data
points, error bars are smaller than symbol size.
}
\label{fig:size_dilute}
\end{figure}

In Fig.~\ref{fig:size_dilute}, simulations results for the re-scaled mean-square
end-to-end distance of bottlebrushes, $\langle R_0^2\rangle/N_{\mbox{\tiny bb}}^{6/5}$, are plotted as
a function of the scaling variable, $(1+N_{\mbox{\tiny sc}}z)^{4/5}$.
Our numerical data corroborate the scaling prediction of Eq.~(\ref{rsd.eq}).
The extension of the bottlebrush backbone caused by the increase of
$N_{\mbox{\tiny sc}}$ and $z$ is due to the enhancement of steric repulsion between densely grafted side chains.
The data in Fig.~\ref{fig:size_dilute} also indicate
that the mean-square size levels off for bottlebrushes whose backbones are too short.
This is attributed to the finite extensibility of the
backbone, which becomes more relevant with increasing values of $N_{\mbox{\tiny sc}}$
and {\em decreasing} values of $z$. At first glance, the latter
observation seems counter-intuitive, because one may assume that stretching of
the backbone becomes more pronounced as both the number of monomers per side
chains  and the grafting density  increase.
Our theoretical approach
reflects that decreasing the grafting density leads
to a stronger effect of finite extensibility.
The backbone remains semi-flexible, if the persistence length
is sufficiently smaller than the overall extension of the
macromolecule, i.e., $R_0/l\gg 1$. With Eqs.~(\ref{n.eq}) and (\ref{rsd.eq}),
one obtains the condition $N_{\mbox{\tiny bb}}\gg\sqrt{N_{\mbox{\tiny sc}}/z}$, which is in qualitative accordance with our observation from Fig.~\ref{fig:size_dilute}.
Note that the effect of finite extensibility diminishes as the concentration is
increased. For bottlebrushes with $N_{\mbox{\tiny bb}}\geq100$, it has been shown that
finite extensibility can be neglected under melt conditions \cite{paturej_sci}.
In what follows, we restrict our analysis to the case of bottlebrushes with $N_{\mbox{\tiny bb}}=100$.

\subsection*{Concentration-dependence of bottlebrush solutions}
\begin{figure}[!h]
\center
\includegraphics[scale=0.3]{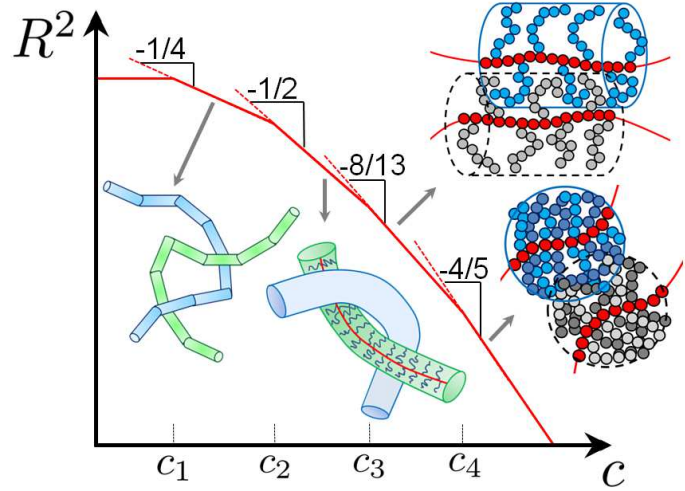}
\caption{{\bf Hierarchical screening of excluded volume in solutions of bottlebrushes.}
  Dependence  of mean-square end-to-end distance, $R^2$,  on  concentration, $c$, for bottlebrush solutions, as predicted theoretically. Bottlebrushes undergo a hierarchical screening of excluded volume interactions
   at different length scales, which results in a sequence of decays in macromolecular size with characteristic exponents as indicated.
   The screening first takes place along the
  backbone of the macromolecule (regime 1 for $c_1<c< c_2$), then along the persistence
  segment (regime 2 for $c_2<c< c_3$), between side chains of different molecules (regime 3 for $c_3<c< c_4$), and between side chains of the same molecule (regime 4 for $c>c_4$).
The scaling predictions for the mean-square size of bottlebrushes in different concentration regimes  and the corresponding overlap concentrations are listed in Table~\ref{table}.
For detailed calculations, see Methods section.
}
\label{fig:dos}
\end{figure}

To study the dependence of conformational properties of
   bottlebrushes on concentration we incorporate the model of a bottlebrush in dilute solution, developed in the previous section,
   and perform scaling analysis (for detailed calculations, see Methods section).
   We predict a sequence of four
regimes  upon increasing concentration, as depicted in Fig.~\ref{fig:dos}. For concentrations $c$ with $c_1<c< c_2$, macromolecules start to overlap
and the excluded volume interaction is screened on the length scale
comparable to the extension of the backbone.
The screening results in a change of conformation from a self-avoiding walk of persistent segments into a random walk of these segments.
This regime is similar to that
for linear chains (without side chains), and the concentration
dependence is the same, i.e., the mean-square end-to-end distance of the macromolecule
scales as $R_1^2\propto c^{-1/4}$~\cite{degennes}. 
\begin{table}[ht]
\def\arraystretch{0.5}
 \begin{center}
 \begin{tabular}{c||c|c|c|c|c|c}
$x$ & $c_x$ & Eq. &$R_x^2$  & Eq.\\ 
\hline\hline
$0$ & -- & -- &$N_{\mbox{\tiny bb}}^{6/5}(N_{\mbox{\tiny sc}}z)^{4/5}c^{0}$  & (\ref{rsd.eq}) \\
$1$ & $N_{\mbox{\tiny bb}}^{-4/5}(N_{\mbox{\tiny sc}}z)^{-1/5}$&(\ref{c1.eq}) & $N_{\mbox{\tiny bb}}(N_{\mbox{\tiny sc}}z)^{3/4}c^{-1/4}$ & (\ref{r1})  \\
$2$ & $N_{\mbox{\tiny sc}}^{-3/5}z^{1/5}$& (\ref{c2.eq})& $N_{\mbox{\tiny bb}}(N_{\mbox{\tiny sc}}z)^{4/5}c^{-1/2}$  &(\ref{r2}) \\
$3$ & $N_{\mbox{\tiny sc}}^{-13/20}z^{11/20}$ &(\ref{c3.eq}) & $N_{\mbox{\tiny bb}}N_{\mbox{\tiny sc}}^{1/2}z^{33/26}c^{-8/13}$  &(\ref{R3a.eq})  \\
$4$ & $N_{\mbox{\tiny sc}}^{-1/2}z^{7/26}$  & (\ref{c4.eq})& $N_{\mbox{\tiny bb}}N_{\mbox{\tiny sc}}^{1/2}z^{157/130}c^{-4/5}$  &(\ref{R4}) \\
\end{tabular}
\end{center}
\caption{Summary of the theoretically predicted overlap concentrations, $c_x$
  ($x=0,\ldots,4$), and  the corresponding mean-square end-to-end
  distances, $R_x^2$, of bottlebrush macromolecules as functions of the degrees of
  polymerization of the backbone ($N_{\mbox{\tiny bb}}$) and the side chains ($N_{\mbox{\tiny sc}}$),
  grafting density of the side chains ($z$), and concentration ($c$). 
}
\label{table}
\end{table}
For larger concentrations, $c_2< c < c_3$, the
screening of excluded volume takes place on a length scale comparable
to the persistence length.
As a consequence,
the backbone monomers of the persistent segment now perform a random walk instead of self-avoiding walk [cf.~Eq.~(\ref{lsaw})].
In this regime, we predict $R_2^2\propto c^{-1/2}$.
Once the concentration is increased even further, such that $c_3<c< c_4$,
side chains of different molecules start to screen each other, which yields a change of statistics for the side chains from self-avoiding walk
to random walk.
This leads to
$R_3^2\propto c^{-8/13}$.
Finally, for highly concentrated  solutions  with $c>c_4$, the
individual macromolecules are strongly compressed and intra-side chain
screening occurs, where side chains of the same macromolecule can screen
their excluded volume interaction.
For this regime, we predict $R_4^2\propto c^{-4/5}$.
The different overlap concentrations and
predictions for the macromolecular size of bottlebrushes in the corresponding regime of
concentrations are summarized  in Table~\ref{table}.
Detailed calculations can be found in the Methods section.

\begin{figure}[!h]
\center
\includegraphics[scale=0.2]{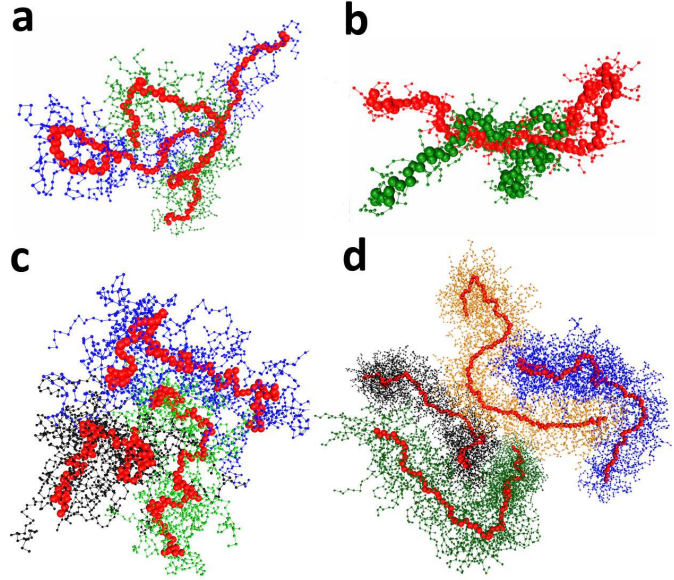}
\caption{{\bf Bottlebrush macromolecules in solutions at different concentrations.}
Simulation snapshots for bottlebrushes of different architecture displaying  hierarchical
    screening of excluded volume interactions, which is observed at different regimes of concentration:
 ({\bf a}) regime 1 ($c_1\!<c\!<c_2$) -- screening along the backbone ($N_{\mbox{\tiny sc}}=8$, $z=1$, $c=0.2$), ({\bf b}) regime 2 ($c_2\!<c\!<c_3$) -- screening along persistent segments ($N_{\mbox{\tiny
    sc}}=4$, $z=1$, $c=0.7$), ({\bf c}) regime 3 ($c_3\!<c\!<c_4$) -- intermolecular screening of the side chains  ($N_{\mbox{\tiny sc}}=32$, $z=1/3$, $c=0.8$),
({\bf d}) regime 4 ($c\!>\!c_4$) -- inter- and intra-molecular screening of the side chains ($N_{\mbox{\tiny
        sc}}=16$, $z=2$, $c=0.8$).
        All macromolecules have the same degree of polymerization of the backbone, $N_{\mbox{\tiny bb}}= 100$,  but different
degrees of polymerization of the side chains, $N_{\mbox{\tiny sc}}$, and grafting density, $z$, as indicated in the parenthesis above.
}
\label{fig:snap}
\end{figure}

 The hierarchy of excluded volume screening
is visualized by MD snapshots
in Fig.~\ref{fig:snap}.
Here, we display bottlebrush conformations representing screening
on different length scales
for all theoretically derived concentration regimes
 ranging from semidilute to highly
concentrated solutions.

\begin{figure}[!h]
\center
\includegraphics[scale=0.3]{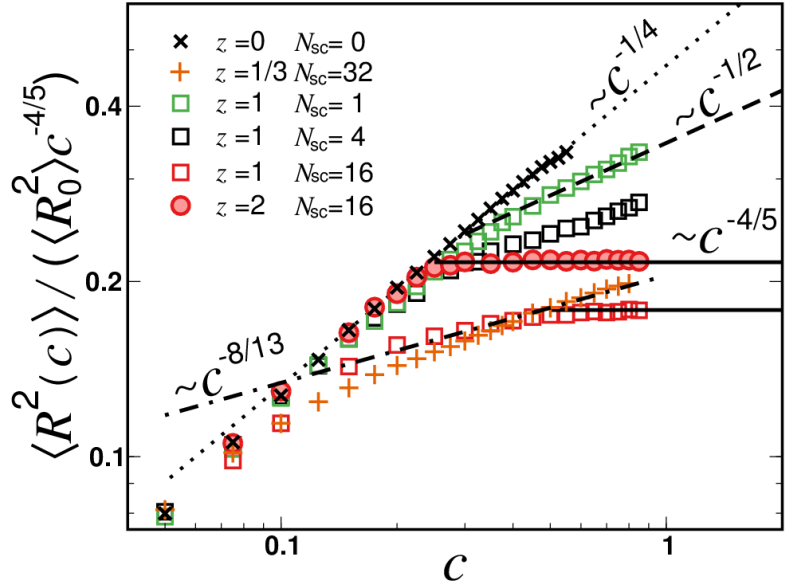}
\caption{{\bf Concentration-dependent size of bottlebrushes.} Mean-square end-to-end distance, $\langle R^2(c)\rangle$, of bottlebrushes at concentration $c$ normalized by
  their values at dilute concentration, $\langle R_0^2\rangle$, and the
  theoretical scaling with concentration in a melt (regime 4), $c^{-4/5}$, as a function
  of  $c$.
  Lines represent theoretical scaling laws for the mean-square size of macromolecules in the corresponding regime of concentration: (dotted)  regime 1 with $R^2_1\propto c^{-1/4}$, (dashed) regime 2 with $R^2_2\propto c^{-1/2}$, (dotted-dashed) regime 3 with $R^2_3\propto c^{-8/13}$, and (solid) regime 4 with $R^2_4\propto c^{-4/5}$.
  Data are shown for
  macromolecules with various degrees of polymerization of side chains, $N_{\mbox{\tiny sc}}$, and grafting
  densities of side chains, $z$, as indicated in
  the legend. The degree of polymerization of the backbone is fixed to $N_{\mbox{\tiny bb}}=100$
  for all data.
}
\label{fig:Rvsc}
\end{figure}

Our simulations confirm the scaling predictions for the concentration-dependent size of bottlebrushes.
The mean-square end-to-end distance, $\langle R^2(c)\rangle$, at concentration $c$ for macromolecules of different architectures
is plotted as a function of $c$ in Fig.~\ref{fig:Rvsc}.
The concentration regime 1 (cf.~Fig.~\ref{fig:snap}a), with scaling $R^2_1\propto c^{-1/4}$ (dotted line), is observed
independently of the degree of polymerization of side chains  and grafting density.
In particular, linear chains ($z=0$) obey the well-known scaling result
in the whole range of considered concentrations with $c>c_1$.
 The dependence of size on concentration for crew-cut bottlebrushes with $N_{\mbox{\tiny sc}}\lesssim 4$ (cf.~Fig.~\ref{fig:snap}b) is well described by  the scaling law $R_2^2\propto c^{-1/2}$  in regime 2  (dashed line).
\begin{figure}[!h]
\center
\includegraphics[scale=0.3]{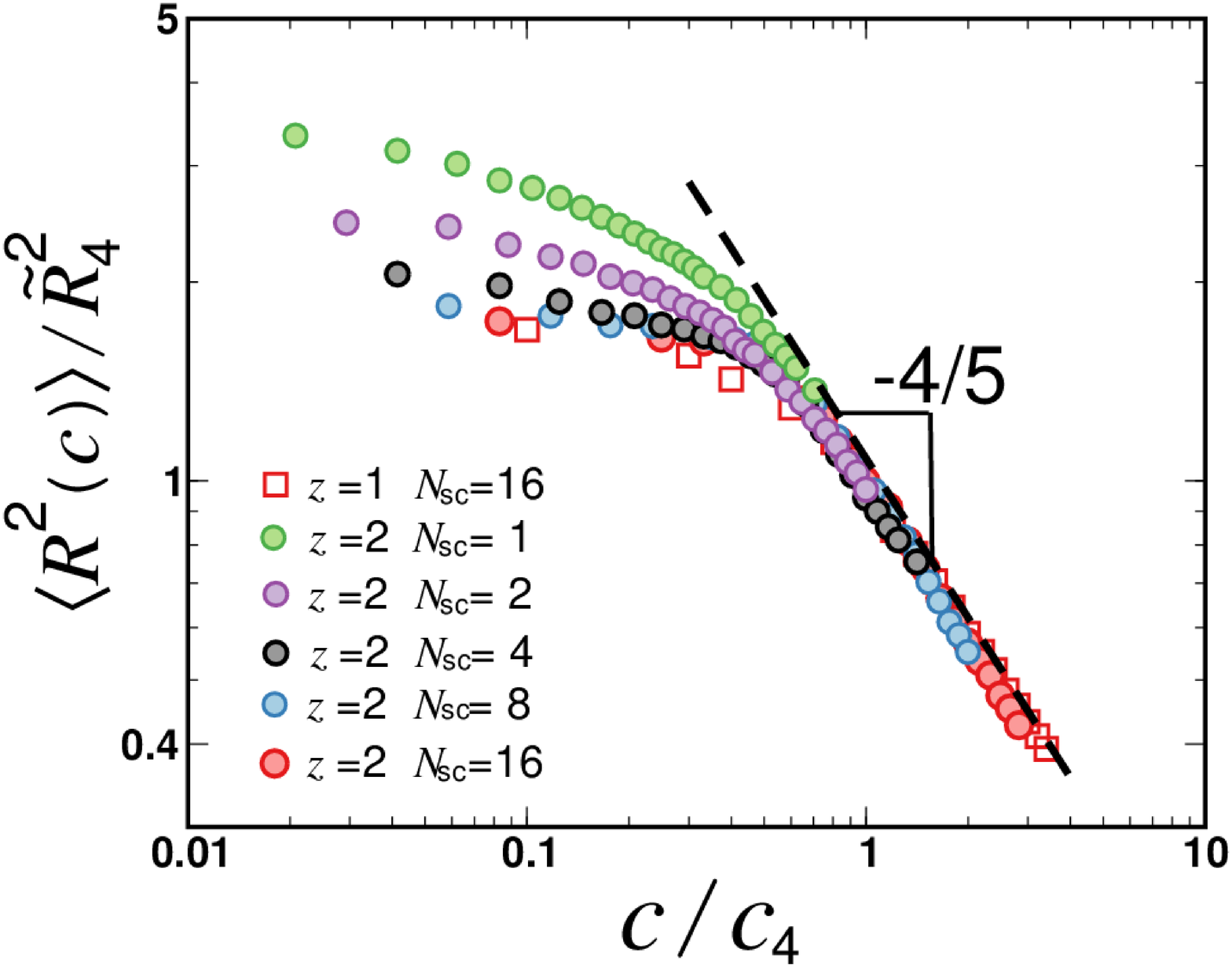}
\caption{{\bf Universal scaling plot of concentration-dependent size of bottlebrushes.}
Double-logarithmic scaling plot for bottlebrush extension versus concentration.
The mean-square end-to-end distance, $\langle R^2(c) \rangle$,
 of bottlebrushes at concentration $c$
is normalized by its value,  $\tilde{R}^2_4\propto N_{\mbox{\tiny bb}}N_{\mbox{\tiny sc}}^{9/10}z^{129/130}$,
at the overlap concentration $c_4$ for regime 4 (melt)
   as a function of $c$ normalized by 
   $c_4\propto N_{\mbox{\tiny sc}}^{-1/2}z^{7/26}$.
The dashed line represents the theoretical scaling law for regime 4 (see Eq.~\ref{r4scal.eq}).
   Data are for
  macromolecules with various degrees of polymerization of side chains, $N_{\mbox{\tiny sc}}$, and grafting
  densities of side chains, $z$, as indicated in
  the legend. The degree of polymerization of the backbone is fixed to $N_{\mbox{\tiny bb}}=100$
  for all data.}
\label{fig:R4}
\end{figure}
This is attributed to negligible screening
 of excluded volume interactions among side chain monomers.
Consequently,  macromolecules with short side chains remain in regime 2 for all concentrations
with $c>c_2$.
The regimes 3 and 4 incorporate screening of excluded volume by mutual side chain interactions and are relevant for macromolecules with long side chains ($N_{\mbox{\tiny sc}}\gtrsim 16$).
For comb-like macromolecules with grafting density $z=1/3$ (cf.~Fig.~\ref{fig:snap}c),
inter-side chain screening occurs and
the data reproduce the scaling in regime 3, $R_3^2\propto c^{-8/13}$
(dotted-dashed line), for concentrated solutions with $c>c_3$.
For bottlebrushes with $z=1$ and $z=2$ in melts (cf.~Fig.~\ref{fig:snap}d), the overlap
between side chains of neighboring molecules is limited \cite{paturej_sci}. Thus, intra-side chain
contacts dominate and are responsible for screening of excluded volume interactions.
The latter conclusion corroborates with the numerical data, which display the scaling
law $R_4^2\propto c^{-4/5}$ of regime 4 (solid lines) for concentrations with $c>c_4$.

The structural properties of highly grafted bottlebrushes in melts, described by regime 4,  are of fundamental importance in shaping mechanical and physical properties of novel polymeric materials.
To perform a universal scaling plot
for this regime, we  derive
[see Methods section, Eq.~(\ref{R4})]
\begin{equation}
\left(\frac{R_4^2}{\tilde{R}_4^2}\right)\propto \left(\frac{c}{c_4}\right)^{-4/5},
\label{r4scal.eq}
\end{equation}
with $\tilde{R}^2_4\propto N_{\mbox{\tiny bb}}N_{\mbox{\tiny sc}}^{9/10}z^{129/130}$ and $c_4\propto N_{\mbox{\tiny sc}}^{-1/2}z^{7/26}$.
We test the prediction of Eq.~(\ref{r4scal.eq})  in Fig.~\ref{fig:R4} and find a
striking agreement with the data from the MD
simulations. Figure~\ref{fig:R4} demonstrates that the universal scaling plot enables superposition of all data
onto a master curve for concentrations with $c>c_4$.
We emphasize that we have not introduced any numerical pre-factors. This means that the proportionality
sign in Eq.~(\ref{r4scal.eq}) can be replaced by an equality sign and all pre-factors
cancel out.
The scaling plot should be  particularly useful for
experimental data, which may become available in the near
future.

In theory, scaling plots like the one displayed in Fig.~\ref{fig:R4} can be constructed for all concentration regimes. However, one needs a clear separation
of the involved length scales, such that the different concentration regimes are wide enough. The width of an individual concentration regime follows from the ratio
of consecutive overlap concentrations, see Table~\ref{table}.  In simulations, the separation of length scales is very difficult to achieve, because one needs bottlebrushes
with relatively large backbones and side chains, which requires significant computational efforts.

\section*{Discussion}

The hierarchical screening of  excluded volume interactions in bottlebrush solutions
stems from a variety of length scales characterizing their architecture
and leads to
distinct scaling regimes for
their conformational properties.
We demonstrated that
the behavior of the concentration-dependent size known for linear chains 
(without grafted side chains)
is reproduced as
long as the size of the backbone  represents the only relevant length scale.
This condition is fulfilled for low concentrations just above the onset of macromolecular overlap.
As the
concentration is increased,
screening emerges
on the substructure of the macromolecule.
First, the persistence length, the length
at which the backbone monomers are correlated due to steric repulsion of the grafted side chains,
becomes the next lower relevant length scale.
Upon further increase of the concentration,
the length of grafted side chains comes into play.
Here,
we observe the consequences of
excluded volume screening along the interacting side chains of different macromolecules. For melts
  of bottlebrushes with  sufficiently long side chains, an ultimate regime appears
due to screening of excluded volume among side chains within the
same macromolecule.

Our theoretical approach differs  from the previous model by Borisov {\em et al.}~\cite{borisov},
 but we obtain a very similar scaling law in the regime of concentrated solutions (regime 3), where inter-side chain screening dominates conformational
properties.
 However, the concentration regime that is characterized by
intra-molecular side chain screening (regime 4) has not been anticipated before.
Our MD simulation provides a significant and successful verification of this novel regime.
The available
 results from scattering experiments
\cite{boli} indicate a decrease in bottlebrush size with increasing
concentration, but an accurate comparison to scaling predictions is not possible yet, due to the limited number of data points.
Therefore, we believe that our results will stimulate new experimental investigations.

Our study demonstrates that conformations of branched macromolecules display hierarchical transitions facilitated by screening
of excluded volume interactions on various length scales.
This phenomenon is
generic in nature and does not depend
on the specific arrangement of grafting nodes.
A broad range of polymeric architectures, including dendromers, dendronized, pom-pom-shaped and star-burst polymers,  are expected to demonstrate
similar
hierarchical screening effects.
Therefore, our work presents valuable insights into structural properties of topologically complex polymers in solutions and melts.

\section*{Materials and Methods}
\subsection*{Molecular dynamics simulations}
Simulations of bottlebrush, comb-like, and linear polymer solutions are performed
using the coarse-grained bead-spring model of Kremer and Grest \cite{kg}. An
individual macromolecule is composed of $N_{\mbox{\tiny bb}}$ backbone
monomers (modeled as excluded volume spheres), which are connected by
bonds. To these backbone monomers, we connect side chains with grafting density
$z$. Each side chain contains $N_{\mbox{\tiny sc}}$ monomers, which are identical to the monomers of the backbone.
The total number of beads in a molecule is $N_{\mbox{\tiny bb}} (1+  N_{\mbox{\tiny sc}}z)$.

The non-bonded interactions between monomers separated by a distance $r$ are
modeled by the truncated and shifted Lennard-Jones (LJ) potential,
 \begin{equation}
 V^{\mbox{\tiny LJ}}(r) =\left\{ {\begin{array}{*{20}c}
   4\epsilon\left[
(\sigma/ r)^{12} - (\sigma /r)^6 + C\right]  \qquad r \leq r_c \\
   0~\qquad\qquad\qquad\qquad\qquad\qquad r > r_c, \\
\end{array}} \right.
\label{wca}
\end{equation}
where the interaction strength, $\epsilon$, is measured in units of thermal
energy, $k_{\rm B}T$, $\sigma$ is the monomer diameter, $r_c$ is the cutoff, and $C$
is the shift of the potential introduced to avoid a discontinuity at $r=r_c$.
We  use $\epsilon=k_{\rm B}T$, $C=1/4$, and  $r_c=2^{1/6}\,\sigma$.
This choice of parameters
results in purely repulsive interactions between monomers, ensuring good
solvent conditions at all values of $k_{\rm B}T$.

The bonded interactions in a molecule are mimicked by the Kremer-Grest potential \cite{kg},
$V^{\mbox{\tiny KG}}(r)=V^{\mbox{\tiny FENE}}(r) + V^{\mbox{\tiny LJ}}(r)$, with the
 ``finitely extensible nonlinear elastic'' (FENE) potential
\begin{equation}
V^{\mbox{\tiny FENE}}= -\frac 12 k r_{\mbox{\tiny F}}^2 \ln{\left[ 1 - \left(\frac
r{r_{\mbox{\tiny F}}}\right)^2 \right]}.
\label{fene}
\end{equation}
Here, the bond spring-constant is $k=30\,\epsilon/\sigma^2$, and the maximum
bond length is $r_{\mbox{\tiny F}}=1.5\,\sigma$ \cite{kg}.
All simulations are performed in a cubic box
with periodic boundary conditions imposed in all spatial dimensions.

We use the Velocity-Verlet algorithm \cite{verlet} to solve the Langevin equation of motion for
the position $\mathbf r_i$ of each monomer with mass $m$,
\begin{equation}
m\ddot{\mathbf r}_i =   \mathbf F_i^{\mbox{\tiny
LJ}} + \mathbf F_i^{\mbox{\tiny FENE}} -\zeta\dot{\mathbf r}_i + \mathbf F_i^{\mbox{\tiny R}}.
\label{langevin}
\end{equation}
The forces $\mathbf F_i^{\mbox{\tiny LJ}}$ and $\mathbf F_i^{\mbox{\tiny
    FENE}}$ respectively follow from the LJ (Eq.~\ref{wca}) and the FENE (Eq.~\ref{fene}) interaction potentials.
The third and fourth term on the right hand side of Eq.~(\ref{langevin}) are a slowly evolving viscous force, $-\zeta\dot{\mathbf r}_i$,
and a rapidly fluctuating stochastic force, $\mathbf
F_i^{\mbox{\tiny R}}$.  The random force, $\mathbf
F_i^{\mbox{\tiny R}}$, is related to the friction coefficient, $\zeta$, by the fluctuation-dissipation
theorem, $\langle \mathbf F_i^{\mbox{\tiny R}}(t) \mathbf F_j^{\mbox{\tiny R}}(t')\rangle = k_{\rm B}T\zeta \delta_{ij}\delta(t-t')$.
The friction coefficient used in our simulations is $\zeta=0.5\,m\tau^{-1}$,
where $\tau = \sqrt{m\sigma^2/\epsilon}$ is the LJ time unit.
 The integration step is taken to be $\Delta \tau = 0.005\tau$, and the
 thermal energy is constant at $k_{\rm B}T=1$.
 All simulations are carried out using the Large-scale Atomic/Molecular
 Massively Parallel Simulator (LAMMPS) \cite{lammps}, and the simulation
 snapshots are rendered using the program Visual Molecular Dynamics (VMD) \cite{vmd}.
Initially, molecules are grown using a self-avoiding random walk technique and placed randomly in the simulation cell.
The initial concentration of all systems is small, $c\approx 5\cdot10^{-4}\, \sigma^{-3}$.
To obtain the desired concentration, the simulation box is gradually decreased in size 
at constant velocity $10^{-3}\, \sigma/\tau$. Once the target density is reached, simulations are continued for up to at least
three relaxation times of the corresponding system.
During the equilibration stage, the molecules diffuse on average at least over the  root-mean-square  end-to-end distance of their backbones.

Simulations of solutions are carried out for linear ($z=0$), comb-like ($z=1/3$), and  bottlebrush $(z\geq 1)$ polymers
 for a fixed number of backbone monomers,  $N_{\mbox{\tiny bb}}=100$, in the range of concentration varied
from  $c=0.001\, \sigma^{-3}$ (dilute solutions) to  $c=0.85\, \sigma^{-3}$ (melts).
The number of monomers  per side chain is varied between $N_{\mbox{\tiny sc}}=0$ and $N_{\mbox{\tiny sc}}=16$ for bottlebrushes with $z=1$ or $z=2$
side chains attached to each backbone monomer. For macromolecules with $z=1/3$, the number of side chains
monomers is fixed to $N_{\mbox{\tiny sc}}=32$.
In addition, for dilute solutions of bottlebrushes ($c=0.001\,\sigma^{-3}$), the number of backbone monomers is varied ($N_{\mbox{\tiny bb}}$=50, 100, and 200) as
well as the number of side chain monomers ($N_{\mbox{\tiny sc}}=1$, 2, 4, 8, 16, 32, and 64) and the grafting density
($z=1$, 2, 3, and 4).

\subsection*{Scaling analysis}
Before presenting our scaling analysis for the concentration-dependent  properties
of bottlebrush conformations, we would like to discuss briefly how our approach
compares to the previous models by
Birshtein {\it et al.}~\cite{bir}
and Fredrickson \cite{fred},
which start out from rod-like backbones, at least, on a
local scale. To do so, we revisit our starting point, the free energy of the
 cylindrical subsegment as given by Eq.~(\ref{F.eq}).
 Instead of
minimizing the free energy with respect to $l_0$ and $R_{\mbox{\tiny sc},0}$, one may
assume a stiff backbone inside the cylinder, i.e., $l_0\propto n_0$. Minimization
of $F$ with respect to $R_{\mbox{\tiny sc},0}$ then yields
\begin{equation}
\label{ws.eq}
R_{\mbox{\tiny sc,0}}\propto N_{\mbox{\tiny sc}}^{3/4}z^{1/4}.
\end{equation}
The above scaling law for the size of side chains is known for bottlebrushes with rod-like backbones, see
Refs.~\cite{bir,fred,wang}.

The spatial distance between grafting points for bottlebrushes with rod-like backbones, $d$, may
be derived from a simple scaling approach,
\begin{equation}
 R_{\mbox{\tiny sc,0}}=\tilde{g}(r_0/d)\propto N_{\mbox{\tiny sc}}^{3/4}z^{1/4},
\end{equation}
where $r_0\propto N^{3/5}$ denotes the size of a linear (not grafted) chain in
dilute solution, and $\tilde{g}(x)$ is a scaling function. This ansatz leads to
\begin{equation}
d\propto z^{-1}
\label{dstiff}
\end{equation}
and, thus, to a rigid
backbone inside the cylinder.
The very same idea can be applied to the side chain scaling derived in the
main text for bottlebrushes with semi-flexible backbones [see Eq.~(\ref{r.eq})  in the limit $N_{\mbox{\tiny sc}}z \gg 1$
, i.e., $R_{\mbox{\tiny sc}}\propto N_{\rm sc}^{7/10}z^{1/10}$].
The scaling argument yields
\begin{equation}
d\propto z^{-3/5},
\label{dsaw}
\end{equation}
 such that the spatial distance between grafting points
resembles a self-avoiding walk statistics \cite{degennes}.
This is the fundamental difference between the scaling law of Eq.~(\ref{ws.eq}) and our result, cf.~Eq.~(\ref{r.eq}).

In the following, we present our scaling analysis for concentrations ranging from semi-dilute solutions to melts.
The macromolecules start to overlap as the
concentration is increased above their overlap concentration, $c>c_1$. With
Eq.~(\ref{rsd.eq}), the overlap concentration reads
\begin{equation}
\label{c1.eq}
c_1\propto  \frac{N_{\mbox{\tiny bb}}(1+N_{\mbox{\tiny sc}}z)}{R_0^3}\propto N_{\mbox{\tiny bb}}^{-4/5}(1+N_{\mbox{\tiny sc}}z)^{-1/5}.
\end{equation}
In the semi-dilute regime, the screening of excluded volume interactions
along the backbones
 is due to the presence
of other macromolecules
and leads to a random walk of the persistence segments \cite{degennes}.
 Therefore, the size of a bottlebrush  scales with the number of backbone monomers as
\begin{equation}
R_1\propto R_0\tilde{g}_1(c/c_1)\propto N_{\mbox{\tiny bb}}^{1/2},
\end{equation}
where $\tilde{g}_1(x)$ is a scaling function.
Together with Eqs.~(\ref{rsd.eq}) and (\ref{c1.eq}), we obtain the size of a bottlebrush in concentration regime 1, which reads
\begin{equation}
R_1\propto (1+N_{\mbox{\tiny sc}}z)^{3/8}N_{\mbox{\tiny bb}}^{1/2}c^{-1/8}.
\label{r1}
\end{equation}
Note that for  macromolecules with $N_{\mbox{\tiny sc}}z=0$, Eq.~(\ref{r1})
 reproduces the expected power law  dependence for linear chains,
$R_{\mbox{\tiny linear}}\propto N_{\mbox{\tiny bb}}^{1/2}c^{-1/8}$
\cite{degennes}, which has been confirmed experimentally \cite{daoud} and by computer simulations \cite{cstarkremer,cstar2}.

Upon further increase of concentration, $c>c_2$, the persistence segments of neighboring bottlebrushes
start to overlap. The corresponding overlap concentration reads
\begin{equation}
\label{c2.eq}
c_2\propto\frac{n_0(1+N_{\mbox{\tiny sc}}z)}{l_0R^2_{\mbox{\tiny sc},0}}\propto
\left(\frac{N_{\mbox{\tiny sc}}}{z}\right)^{-2/5}(1+N_{\mbox{\tiny sc}}z)^{-1/5},
\end{equation}
where we have used Eqs.~(\ref{n.eq}) and (\ref{lsaw}).
In concentration regime 2, the
self-avoiding walk of monomers inside the cylinder turns into a random
walk. Since the excluded volume contribution of the side
chains remains unaltered, one may anticipate that the persistence length in this regime is given by
\begin{equation}
\label{l2.eq}
l_2\propto n_2^{1/2}(1+N_{\mbox{\tiny sc}}z)^{2/5}.
\end{equation}

A priori, we do not know how the number of backbone monomers inside the
cylinder for regime 2, $n_2$, depends on $N_{\mbox{\tiny sc}}$ and
$z$. However, we may assume that $n_2$ remains proportional to $z^{-1/2}$, see
Eq.~(\ref{n.eq}). Thus, in the limit $N_{\mbox{\tiny sc}}z\gg 1$,
Eq.~(\ref{l2.eq}) suggests  $l_2\propto z^{3/20}$. The latter
result allows us to perform a crossover scaling,
\begin{equation}
 l_2\propto l_0 \tilde{g}_2(c/c_2)\propto z^{3/20},
 \label{l_2}
\end{equation}
with $\tilde{g}_2(x)$ a scaling function.
Together with Eq.~(\ref{lsaw}) and $N_{\mbox{\tiny sc}}z\gg 1$,
Eq.~(\ref{l_2}) leads to
\begin{equation}
l_2\propto R_{{\mbox{\tiny sc}},2}\propto N_{\mbox{\tiny sc}}^{11/20}z^{3/20}c^{-1/4},
\label{l2f}
\end{equation}
where $R_{{\mbox{\tiny sc}},2}$ denotes the size of  side chains in
regime 2.
Within our theoretical picture, both persistence length and size of side chains
depend on concentration, but the number of backbone monomers per cylindrical segment in a
given concentration regime does not. Thus, Eq.~(\ref{l2f}) suggests a
decreasing persistence (side chain) length with a constant number of backbone
monomers inside the persistence segment.

Equation (\ref{l2f}) can be re-written as $l_2\propto N_{\mbox{\tiny
    sc}}^{3/20}z^{-1/4}(N_{\mbox{\tiny sc}}z)^{2/5}c^{-1/4}$. Together with  Eq.~(\ref{l2.eq}) and $N_{\mbox{\tiny sc}}z\gg 1$, we obtain
\begin{equation}
n_2\propto \sqrt{\frac{N_{\mbox{\tiny sc}}^{3/5}}{z}},
\label{n_2}
\end{equation}
which reflects a natural modification of Eq.~(\ref{n.eq}).
With Eqs.~(\ref{l2f}) and (\ref{n_2}),  we obtain the macromolecular size of bottlebrushes in
regime 2 ($N_{\mbox{\tiny sc}}z\gg 1$),
\begin{equation}
R_2\propto (1+N_{\mbox{\tiny sc}}z)^{2/5}N_{\mbox{\tiny bb}}^{1/2}c^{-1/4}.
\label{r2}
\end{equation}

Once the concentration is increased even further, $c>c_3$, the macromolecules
attain melt concentration and side chains of neighboring macromolecules start to
overlap. The corresponding overlap concentration reads
\begin{equation}
\label{c3.eq}
c_3\propto \frac{N_{\mbox{\tiny sc}}z}{l_2R_{{\mbox{\tiny sc}},2}^2}\propto N_{\mbox{\tiny sc}}^{-13/20}z^{11/20},
\end{equation}
where Eq.~(\ref{l2f}) in the limit  $N_{\mbox{\tiny sc}}z\gg 1$ has been used.
Due to the screening of excluded volume interactions along the side chains of neighboring macromolecules,
one expects that the size of side chains, $R_{{\mbox{\tiny sc}},3}$, and the corresponding persistence length, $l_3$, in regime 3 scale as $R_{{\mbox{\tiny sc}},3}\propto l_3\propto N_{\mbox{\tiny sc}}^{1/2}$.
With $\tilde{g}_3(x)$ a scaling function, the crossover scaling
\begin{equation}
l_3\propto l_0\tilde{g}_3(c/c_3)\propto N_{\mbox{\tiny sc}}^{1/2}
\end{equation}
leads to
\begin{equation}
\label{l3.eq}
l_3\propto R_{{\mbox{\tiny sc}},3} \propto N_{\mbox{\tiny sc}}^{1/2}z^{7/26}c^{-4/13},
\end{equation}
where we have used Eqs.~(\ref{lsaw}) and (\ref{c3.eq}) in the limit $N_{\mbox{\tiny sc}}z\gg 1$.
In highly concentrated solutions, the persistence length is
proportional to the number of side chains within the cylinder, i.e.,
$l_3\propto n_3z$. With Eq.~(\ref{l3.eq}), the number of backbone monomers in
the cylinder  then is
\begin{equation}
n_3\propto N_{\mbox{\tiny sc}}^{1/2}z^{-19/26}.
\end{equation}
The size of bottlebrushes  in regime 3, $R_3$, follows from a random walk of persistence segments with length $l_3$, where  each  segment contains $n_3$ backbone monomers.
This yields
\begin{equation}
R_3\propto l_3\left(\frac{N_{\mbox{\tiny bb}}}{n_3}\right)^{1/2}\propto
N_{\mbox{\tiny bb}}^{1/2}N_{\mbox{\tiny sc}}^{1/4}z^{33/52}c^{-4/13}.
\label{R3a.eq}
\end{equation}
The above scaling result 
is very close to the one  predicted 
by Borisov {\it et al.}~\cite{borisov} with respect to all four exponents.
However, the underlying assumptions for both models are different.
The scaling of bottlebrush size with $N_{\mbox{\tiny bb}}$ and $N_{\mbox{\tiny sc}}$ has been  confirmed recently  under melt conditions \cite{paturej_sci}.

With respect to suppressing entanglement effects in order to design
super-elastic rubbers, highly grafted bottlebrushes  are of
particular interest. Here, an additional regime can appear, where compression of the backbone and the side chains can lead
to mutual screening of side chains that belong to the same macromolecule.
 With Eq.~(\ref{l3.eq}), the overlap concentration in regime 4 is
\begin{equation}
c_4\propto \frac{N_{\mbox{\tiny sc}}z}{R_{{\mbox{\tiny sc}},3}^3}\propto N_{\mbox{\tiny sc}}^{-1/2}z^{7/26}.
\label{c4.eq}
\end{equation}
The crossover scaling reads
\begin{equation}
l_4\propto l_0\tilde{g}_4(c/c_4)\propto N_{\mbox{\tiny sc}}^{1/2},
\end{equation}
with $\tilde{g}_4(x)$ a scaling function.
Using Eq.~(\ref{lsaw}) in the limit $N_{\mbox{\tiny sc}}z\gg 1$, one obtains the concentration dependence of the persistence length  in regime 4,
\begin{equation}
l_4\propto N_{\mbox{\tiny sc}}^{1/2}z^{27/130}c^{-2/5}.
\label{l4.}
\end{equation}
Once more, we assume local stretching of the backbone, i.e., $l_4\propto n_4z$, where $n_4$ denotes the number of backbone
monomers per persistent segment for regime 4. With Eq.~(\ref{l4.}), this leads to
\begin{equation}
R_4\propto l_4\left(\frac{N_{\mbox{\tiny bb}}}{n_4}\right)^{1/2}\propto N_{\mbox{\tiny bb}}^{1/2}N_{\mbox{\tiny sc}}^{1/4}z^{157/260}c^{-2/5},
\label{R4}
\end{equation}
for the size of bottlebrushes in regime 4.
Together with Eq.~(\ref{c4.eq}),
the above equation can be rewritten, such that we obtain Eq.~(\ref{r4scal.eq}) of the main text.

\vspace*{1cm}
\setstretch{1}
\noindent {\bf Acknowledgments:}\\ The authors thank
A. Johner and J.-U. Sommer  for fruitful discussions.  J.P. thanks for computational time
at PL-Grid (Poland) and ZIH (Germany) infrastructures. \\
\noindent \textbf{Funding:}\\  J.P. acknowledges support from the German
Science Foundation (DFG-Pa 2860/2-1)  and the Polish Ministry of Science and Higher Education (IP 2015 059074).
 T.K. thanks the German
Science Foundation (DFG-Kr 2854-2) for financial support. \\

\end{document}